\documentclass[conference]{IEEEtran}

\usepackage{cite}
\usepackage{amsmath,amssymb,amsfonts}
\usepackage{graphicx}
\usepackage{booktabs}
\usepackage{balance}
\usepackage{microtype}
\usepackage{tikz}
\usetikzlibrary{shapes.geometric, arrows.meta, positioning, fit, backgrounds, calc}

\begin{document}

\title{Reimagining Peer Review Process Through Multi-Agent Mechanism Design}

\author{
    \IEEEauthorblockN{Ahmad Farooq}
    \IEEEauthorblockA{\textit{University of Arkansas at Little Rock} \\
    Little Rock, Arkansas, USA \\
    afarooq@ualr.edu \\
    ORCID: 0009-0002-3684-5876}
    \and
    \IEEEauthorblockN{Kamran Iqbal}
    \IEEEauthorblockA{\textit{University of Arkansas at Little Rock} \\
    Little Rock, Arkansas, USA \\
    kxiqbal@ualr.edu \\
    ORCID: 0000-0001-8375-290X}
}

\maketitle

\begin{abstract}
The software engineering research community faces a systemic crisis: peer review is failing under growing submissions, misaligned incentives, and reviewer fatigue. Community surveys reveal that researchers perceive the process as ``broken.'' This position paper argues that these dysfunctions are \emph{mechanism design} failures amenable to computational solutions. We propose modeling the research community as a stochastic multi-agent system and applying multi-agent reinforcement learning to design incentive-compatible protocols. We outline three interventions: a credit-based submission economy, MARL-optimized reviewer assignment, and hybrid verification of review consistency. We present threat models, equity considerations, and phased pilot metrics. This vision charts a research agenda toward sustainable peer review.
\end{abstract}


\section{Introduction}

The ICSE 2026 Future of Software Engineering call opens with a sobering observation: beneath thriving conferences lies ``an undercurrent of grumbling'' suggesting ``our community may not be doing so well.'' Among the complaints, one dominates: ``Peer review is broken.'' ~\footnote{https://conf.researchr.org/track/icse-2026/icse-2026-future-of-software-engineering}

This is not hyperbole. The NeurIPS 2014 duplicate-review consistency experiment found that 57\% of papers accepted by one committee were rejected by the other, highlighting substantial outcome instability~\cite{brezis2020arbitrariness,cortes2021inconsistency}. Studies document that 6.5--16.9\% of AI conference reviews show substantial LLM involvement~\cite{liang2024monitoring}, and that, in LLM-based agent simulations of peer review, up to 37.1\% of variance in decisions is attributable to modeled reviewer biases~\cite{jin2024agentreview}. The ``publish or perish'' culture has created a tragedy of the commons: researchers maximize submissions while minimizing review effort~\cite{horta2024crisis}.

The conventional response treats these as social problems requiring cultural change. This position paper offers a different perspective: \textbf{the peer review crisis is a mechanism design problem}, and mechanism design problems yield to engineering~\cite{roughgarden2010algorithmic}.

We propose reconceptualizing the research community as a \emph{stochastic multi-agent system} where agents pursue individual objectives. Current dysfunction arises because reward structures incentivize individually rational but collectively misaligned behaviors. Multi-agent reinforcement learning (MARL) provides tools to redesign these incentives~\cite{hernandez2019survey, gronauer2022multi}.

\begin{figure}[t]
\centering
\begin{tikzpicture}[
    node distance=0.4cm and 0.4cm,
    box/.style={rectangle, draw, rounded corners, minimum width=2.1cm, minimum height=0.7cm, align=center, font=\scriptsize},
    pillar/.style={rectangle, draw, fill=gray!15, minimum width=2.1cm, minimum height=1.6cm, align=center, font=\scriptsize\bfseries},
    arrow/.style={-{Stealth[length=2mm]}, thick},
    feedback/.style={arrow, dashed, rounded corners}
]


\node[pillar] (credit) {Credit\\Economy};
\node[pillar, right=0.6cm of credit] (assign) {MARL\\Assignment};
\node[pillar, right=0.6cm of assign] (verify) {Hybrid\\Verification};

\node[box, above=0.3cm of credit] (sub) {Submissions};
\node[box, above=0.3cm of assign] (rev) {Reviewers};
\node[box, above=0.3cm of verify] (dec) {Decisions};

\node[box, below=0.3cm of credit] (price) {Price\\Dynamics};
\node[box, below=0.3cm of assign] (const) {Constrained\\RL};
\node[box, below=0.3cm of verify] (arg) {Argumentation\\Analysis};


\draw[arrow] (sub) -- (credit);
\draw[arrow] (rev) -- (assign);
\draw[arrow] (dec) -- (verify);

\draw[arrow] (credit) -- (price);
\draw[arrow] (assign) -- (const);
\draw[arrow] (verify) -- (arg);

\draw[arrow] (credit) -- (assign);
\draw[arrow] (assign) -- (verify);

\draw[feedback] (arg.south) -- ++(0,-0.4) 
    -| node[pos=0.25, below, font=\tiny\itshape] {Reward Signal \& Updates} (price.south);

\end{tikzpicture}
\caption{Three-pillar architecture. Note the feedback loop: review quality verification (Pillar 3) directly informs credit issuance and price dynamics (Pillar 1), creating a closed-loop adaptive system.}
\label{fig:architecture}
\end{figure}
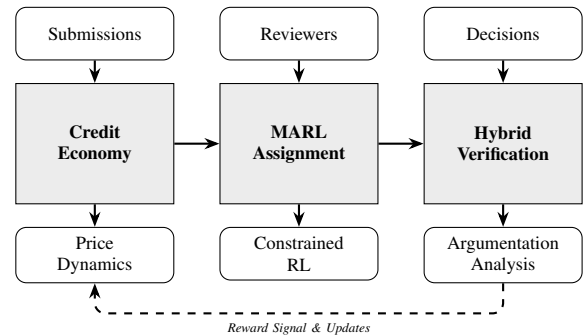

This paper outlines three interventions (Figure~\ref{fig:architecture}) and a phased research agenda. We do not claim these are proven solutions; rather, we argue they constitute a principled framework warranting systematic investigation.

\section{Taking Stock: The Mechanism Failure}

\subsection{The Tragedy of the Review Commons}

The fundamental problem is economic: publishing yields rewards (tenure, promotion), while reviewing yields almost none~\cite{routledge2025improving}. 

This creates a classic commons tragedy. The ``reviewer attention budget'' is finite, but depletion costs are externalized. Overburdened reviewers produce rushed evaluations, leading to ``reviewer roulette''~\cite{liu2023shackles}.

Scientific publishing volume increased by about 47\% between 2016 and 2022, intensifying strain on editorial and reviewer capacity~\cite{hanson2024strain}. Reviewer invitation acceptance rates have declined, so editors may need multiple invitations to secure each completed review~\cite{routledge2025improving}.

\newpage
\subsection{The LLM Amplification Effect}

Large language models are increasingly used in scientific writing~\cite{liang2024mapping} and can reduce the effort required to draft and revise manuscripts. They also enable superficially coherent but substantively hollow reviews. There are growing concerns that LLM assistance may shift review tone and scoring, while detection remains unreliable~\cite{naddaf2025ai}.

\subsection{Related Mechanisms}

Prior work corroborates our approach. Peer prediction methods elicit truthful reports by rewarding agreement with reference raters~\cite{miller2005eliciting}; information-theoretic variants offer stronger incentive guarantees~\cite{kong2019information}. Strategyproof mechanisms~\cite{xu2018strategyproof} provide theoretical rationing guarantees. Isotonic mechanisms use multi-submission knowledge for calibration~\cite{su2021you}.

Recent market-based alternatives like Impact Market~\cite{sankaralingam2025impact} propose decoupling dissemination from credentialing entirely. While promising, our framework seeks to repair the existing workflow rather than replace it, minimizing disruption. Similarly, endogenous matching models have explored linking effort to future assignment probabilities~\cite{xiao2014incentive}. 

Unlike auction-based rationing mechanisms (e.g., VCG combined with peer-prediction~\cite{srinivasan2021auctions}), which clear the market in single-shot rounds, our Credit Economy introduces a persistent, transferable asset. This allows researchers to smooth their labor across time (reviewing now to submit later), addressing the ``bursty'' nature of conference deadlines. Our framework synthesizes these insights, prioritizing practical deployability.

\section{A Multi-Agent Framework}

\subsection{The Community as a Stochastic Game}

We model the research community as a stochastic multi-agent system $\langle \mathcal{A}, \mathcal{S}, \mathcal{T}, \mathcal{R} \rangle$ where agents pursue individual objectives. Currently, $R_{\text{rev}} \approx 0$ while $R_{\text{pub}} > 0$. The solution reshapes rewards so quality reviewing becomes individually rational~\cite{roughgarden2010algorithmic}.

\subsection{Credit-Based Submission Economy}

We propose a ``Review Credit'' (RC) system where submissions cost RC and quality reviews earn RC.

\textbf{Price Dynamics.} Prices follow a closed-loop update rule: $p_{t+1} = p_t + \gamma(D_t - S_t)$ clipped to $[p_{\min}, p_{\max}]$. Here, $D_t$ represents the rolling average of \emph{submission volume} (demand for review slots), and $S_t$ represents the \emph{cleared review capacity} (supply) over epoch $t$. To prevent oscillations (e.g., credit runs), updates occur at fixed monthly epochs with adaptive damping $\gamma$~\cite{kousha2024artificial}. To ensure long-term budget balance and prevent deflationary spirals, the system employs an \emph{adaptive issuance policy}: if the total credit supply drops below a safety threshold (velocity $< V_{\min}$), the protocol temporarily subsidizes review rewards from a central reserve.

\textbf{Stability.} Heterogeneous agent strategies (e.g., hoarding) pose stability risks. We propose monitoring the \emph{Credit Velocity} $V = \frac{\sum \text{Transactions}}{\text{Total Supply}}$ and employing Lyapunov-based control policies to adjust issuance rates if system volatility exceeds safety margins.

\textbf{Supply Policy.} RC issuance occurs through review completion; sinks include submission fees and expiration. Caps prevent hoarding.

\textbf{Quality Measurement.} Review quality combines: blinded author ratings, meta-reviewer consistency, and specificity metrics. Information-theoretic scoring~\cite{kong2019information} offers stronger incentive properties and should be evaluated as a primary method.

\textbf{Newcomer Support.} First-time submitters receive initial endowments; hardship waivers address career breaks; mentoring-linked credit earning provides alternative pathways.

\textbf{Governance Model.} Credits are centrally ledgered by the venue. Disputes are adjudicated by program chairs with escalation to a standing ethics committee; clawbacks require documented misconduct with appeal rights.

\subsection{MARL-Optimized Reviewer Assignment}

Current algorithms optimize primarily for topic match~\cite{jovanovic2023reviewer}. Classical OR methods (min-cost flow, MIP with fairness constraints) excel at static allocation~\cite{charlin2013toronto} but fail to capture dynamic reviewer fatigue.

We propose a Constrained Multi-Agent Reinforcement Learning (CMARL) approach~\cite{achiam2017constrained}.
\begin{itemize}
    \item \textbf{State ($s_t$):} Reviewer load, historical lateness, recent decline patterns, and topic embedding distance.
    \item \textbf{Action ($a_t$):} The assignment matrix subject to hard constraints (COIs, max load).
    \item \textbf{Reward ($r_t$):} A multi-objective function combining timeliness, specificity score, and fairness penalties.
\end{itemize}

\textbf{Counterfactuals.} Offline RL on historical logs suffers from confounding (we only observe outcomes for realized matches). To mitigate this, we propose using \emph{Doubly Robust Estimators} or \emph{Causal Bandits} to estimate the potential reward of counterfactual assignments during the training phase.

\textbf{Learning Setup.} Learning adds value when reviewer behavior is non-stationary or when latent features (past timeliness, decline patterns) improve predictions beyond static matching. Minimum effect sizes (e.g., $>$10\% timeliness improvement) justify added complexity.

\subsection{Hybrid Verification}

Combining structured checklists~\cite{spadini2020primers} with argumentation frameworks~\cite{dung1995acceptability}: reviewers complete rubrics; review text is parsed into claims. 

To ensure scalability, we adopt an ``Agent-as-a-Judge'' paradigm~\cite{zheng2023judging}. An LLM-based verifier extracts claims and checks for stance consistency against the paper content. We acknowledge that LLM verifiers may encode biases or succumb to Goodhart's law (authors writing for the bot). To mitigate this, our pilot employs a ``human-in-the-loop'' audit where a random 10\% of verifications are manually adjudicated, calibrating the agent against expert judgment. Future iterations may also leverage cryptographic watermarking or provenance attestations to verify human authorship, subject to community governance and consent.

Target overhead: under 5 minutes additional time per review.

\section{Threat Model and Mitigations}

Table~\ref{tab:threats} summarizes threats and mitigations for each component.

\begin{table}[t]
\caption{Threat-Mitigation Mapping}
\label{tab:threats}
\scriptsize
\begin{tabular}{@{}p{1.4cm}p{2.0cm}p{3.8cm}@{}}
\toprule
\textbf{Component} & \textbf{Threat} & \textbf{Mitigation} \\
\midrule
Credit Economy & Review rings & Network analysis; randomized audits \\
& Credit hoarding & Expiration; caps; adaptive pricing \\
& Quality gaming & Multi-signal scoring; outlier detection \\
\midrule
Assignment & Bias amplification & Fairness constraints; offline evaluation \\
& Strategic declines & Decline patterns in assignment weights \\
\midrule
Verification & LLM-generated reviews & Disclosure requirements; specificity scoring \\
& Shallow reviews & Argumentation coverage; human triage \\
\midrule
Cross-cutting & Sybil attacks & ORCID + institutional verification \\
& Retaliation & Blinded ratings; temporal smoothing \\
\bottomrule
\end{tabular}
\end{table}

\subsection{Equity and Privacy}

Credit systems risk disadvantaging researchers with fewer opportunities. Fairness safeguards include initial endowments and earning multipliers~\cite{chen2019fairness}. \textbf{Equity Measurement:} Pilots will track credit Gini disaggregated by region, institution type, and career stage; corrective levers address emerging disparities. Privacy-preserving attestations can reconcile auditability with anonymity.

\section{Research Agenda}

\subsection{Phase 0: Simulation \& Foundations (2025--2026)}

Before real-world credit deployment, we will develop an \emph{Agent-Based Model (ABM)} of the conference ecosystem. This simulation will stress-test price stability against strategic behaviors (e.g., collusion rings, free-riding) and calibrate $\gamma$ (damping factors) using historical OpenReview data.

Simultaneously, we deploy lightweight interventions: reviewer accreditation, bi-directional feedback, and structured review templates. These gather empirical baselines and build community acceptance.

\subsection{Phase 1: Pilot Credit System (2026--2027)}

Target venue: ICSE workshop (single-venue, centralized ledger).
\textbf{Methodology:} Randomized Controlled Trial (RCT). To prevent contamination, the \emph{unit of randomization} will be the \emph{Program Committee Area} rather than individual authors. A power analysis will determine the sample size required to detect a 0.5$\sigma$ shift in review timeliness with $\alpha=0.05$.
\textbf{Metrics:} Completion rates ($>$90\%), credit Gini ($<$0.3), newcomer participation.

\subsection{Phase 2: MARL Assignment (2027--2028)}

Shadow mode comparing learning methods against ILP baselines. \textbf{Metrics:} Timeliness ($>$10\% improvement), load balance, fairness across demographics.

\subsection{Phase 3: Verification and Federation (2028--2029)}

Hybrid verification deployment. Workshop pilots expand to co-located conferences, then main venues. Cross-venue federation requires SIG coordination with explicit policies against arbitrage.

\subsection{Governance and Sustainability}

Success requires recognizing review contributions~\cite{aczel2025present}. We recommend sunset clauses, opt-out pathways, transparent audit reports, and annual assessments.

\section{Discussion}

This vision faces challenges. Credit systems can be gamed; learning-based assignments may inadvertently embed biases; and verification risks may incur unnecessary overhead. Agent-based simulations and formal equilibrium analyses are needed before deployment. The mechanisms we propose are starting points, and integration with complementary approaches (information-theoretic scoring, isotonic calibration) merits investigation.

\section{Conclusion}

The peer review crisis arises from misaligned incentives, not moral failure. By applying engineering rigor to governance, we can design mechanisms that make good behavior individually rational. We invite the reviewer community to pursue this research agenda.

\balance
\bibliographystyle{IEEEtran}
\bibliography{output}
\end{document}